%% file: main.tex
\pdfoutput=1

\documentclass[11pt]{article}
\usepackage[numbers]{natbib}
\usepackage[section]{placeins}
\usepackage{xspace}
\usepackage{xcolor}
\definecolor{darkblue}{rgb}{0,0,.5}
\usepackage[colorlinks=true,allcolors=darkblue]{hyperref}
\usepackage{authblk}

\usepackage{algorithm}
\usepackage[noend]{algpseudocode}
\usepackage{bbm}
\usepackage{macros}
\usepackage{tikz}
\usetikzlibrary{matrix}
\usepackage{pgfplots}
\usepackage{overpic}
\usepackage{makecell}
\usepgfplotslibrary{fillbetween}
\usetikzlibrary{patterns}
\usepackage{enumerate}
\usepackage{amssymb}
\usepackage{amsbsy}
\usepackage{amsmath}
\usepackage{amsthm}
\usepackage{amsfonts}
\usepackage{latexsym}
\usepackage{graphicx}
\usepackage{color}
\usepackage{xifthen}
\usepackage{algorithmicx, algpseudocode}
\usepackage{makecell}
\usepackage{cellspace} %
\usepackage{subcaption}
\usepackage{listings}
\newcommand{\N}{\mathbb{N}}

\newtheorem{definition}{Definition}[section]

\usepackage{fullpage}

\newcommand{\printfnsymbol}[1]{%
  \textsuperscript{\@fnsymbol{#1}}%
}

\begin{document}

\title{ 
A Members First Approach to Enabling LinkedIn's Labor Market Insights at Scale}
\author{Ryan Rogers$^*$, Adrian Rivera Cardoso\thanks{Equal contribution} , Koray Mancuhan, Akash Kaura, Nikhil Gahlawat, Neha Jain, Paul Ko, Parvez Ahammad}

\affil{LinkedIn Corporation}
\maketitle

\begin{abstract}
We describe the privatization method used in reporting labor market insights from LinkedIn's Economic Graph, including the differentially private algorithms used to protect member's privacy.  The reports in \url{https://graph.linkedin.com/insights/labor-market} show who are the top employers, as well as what are the top jobs and skills in a given country/region and industry. We hope this data will help governments and citizens track labor market trends during the COVID-19 pandemic while also protecting the privacy of our members. 
\end{abstract}

\input{intro}

\input{prelims}

\input{privReports}

\input{conclusion}

\section{Acknowledgements} Special thanks to the multiple teams that put this project together, including teammates Krati Ahuja, Amar Athavale, Mark Cesar, Sue Duke, Vikram Gaitonde, Aarthi Jayaram, Raviteja Kanumula, Nicholas Kim, Karin Kimbrough. Sofus Macskassy, Jessi Reel, Shraddha Sahay, Steven Shimizu, Venkata Krishnan Sowrirajan, Subbu Subramaniam, Lin Xu, Ya Xu, Sarah Yu, and Zhihao Zhang.

\bibliographystyle{ACM-Reference-Format}
\bibliography{bib}
\end{document}

%% file: intro.tex

\section{Introduction}

The COVID-19 pandemic has turned global labor markets upside down and exacerbated endemic skill and equity gaps. In the wake of the immediate public health crisis, a new normal is emerging for the world of work. Data from LinkedIn's Economic Graph can help governments and citizens track labor market trends and make informed decisions to meet the new future of work as it unfolds.  We provide three different types of metrics that can be sliced by country/region, industry, as well as monthly report date since January 2020.\footnote{Data may be further partitioned at a later date.}  These insights are drawn from LinkedIn's Economic Graph, which is based on global data that includes over 690 million members, 50 million companies, 11 million job openings, 36 thousand skills, and 90 thousand schools.   The three metrics for labor market insights include the following:

\paragraph{Who is hiring?} Understanding which companies are hiring in your area can shed light on industry and market trends, uncover opportunities to get people back to work, and inform workforce development.

\paragraph{What jobs are available?} Understanding what roles are hiring can facilitate more effective investments in economic development initiatives, as well as in the training programs best suited to upskill workers for the jobs available in your region.

\paragraph{What skills are needed?} Understanding the skills needed for the top trending jobs can lay the foundation for investment in workforce development programs to prepare workers in your region for emerging jobs.

\paragraph{} Although we find that this data can be incredibly helpful to governments, policy makers, and individuals in the global workforce during these challenging times, we want to ensure member trust is preserved and no individual can be identified based on the reports we provide.  We then turn to privatization techniques, such as differential privacy \cite{DworkMcNiSm06}, to maintain the privacy of user's data.  Differential privacy provides a way to balance the privacy of users and utility of the aggregated results by injecting carefully calibrated noise into the results.  LinkedIn has deployed privacy systems that utilize differentially private algorithms, including LinkedIn products used by our marketing partners, such as the Audience Engagement API in LinkedIn Marketing Solutions \cite{RogersSuPeDuLeKaSaAh20}.  Other companies and government agencies have developed and deployed privacy systems that utilize differential privacy, including Microsoft \cite{DingKuYe17, Kahan2019, Kahan2020}, Apple \cite{ApplePrivacy17}, Google \cite{ErlingssonPiKo14, GoogleMobility20}, Uber \cite{JohnsonNeSo18}, and the US Census Bureau \cite{DajaniLaSiKiReMaGaDaGrKaKiLeScSeViAb17}.  Recently, Google published mobility reports that show the impact of shelter-in-place mandates due to COVID-19 in different areas using their open source differential privacy library \cite{GoogleMobility20}.

Overall we have 3 types of queries over different dates, industries, and countries/regions, resulting in many reports that must be privatized.  However, due to privacy concerns we only wanted to show reports with large enough counts, hence we show a fewer number of reports. Note that in some reports there are fewer results due to hiring freezes and rising unemployment. This document is intended to go in more technical detail to what private algorithms and parameters were used in releasing these statistics.

We hope these reports can help provide a plan for those seeking to learn new skills in the changing job market, from identifying and learning top skills for the most in-demand jobs, searching and applying for new jobs, and preparing for job interviews to secure job offers.  More detail can be found here: \url{https://opportunity.linkedin.com/}.

%% file: prelims.tex

\section{Preliminaries}

We present notation and fundamental definitions that will be used to describe our privacy approach.  
We will denote the data histogram as $\bbh \in \N^{d}$ where $d$ is the dimension of the data universe, which might be unknown or known.  We say that $\bbh$ and $\bbh'$ are neighbors if they differ in the presence or absence of at most one hire.  Recall the definition of differential privacy \cite{DworkMcNiSm06, DworkKeMcMiNa06}.
\begin{definition}[Differential Privacy]
A randomized algorithm $\cM$ that takes a histogram in $\N^d$ to some arbitrary outcome set $\cY$ is $(\diffp,\delta)$-differentially private (DP) if for all neighbors $\bbh,\bbh'$ and for all outcome sets $S \subseteq \cY$, we have
$
\Pr[\cM(\bbh) \in S] \leq e^\diffp \Pr[\cM(\bbh') \in S] + \delta.
$
If $\delta=0$, then we simply write $\diffp$-DP.
\end{definition} 

The definition of neighboring differs depending on the context or assumptions made and is sometimes referred to as the \emph{granularity} of privacy.  For instance, \emph{event-level} privacy defines neighboring to be the addition or deletion of a single record in the data, despite the actual number of records a user can actually contribute to the dataset being larger.  Alternatively, \emph{user-level} privacy is a more stringent condition that considers additions or deletions of all possible records of a user.  The stronger granularity of privacy can be enforced by either restricting the number of contributions any single user can have in the original dataset or by developing private algorithms that restrict the number of results returned, despite a user contributing an arbitrary number of records in the dataset, see e.g. \cite{DurfeeRo19}.  For this use case we use event-level privacy, where the event is a user being hired.  We note that in our dataset, in any given 3 month period at least 95\% of users are hired at most once.  

The three different metrics we want to provide in the reports are top: jobs, employers, and skills. In particular, we are returning the top-20 results in each metric (or fewer if there are not enough results with sufficient counts).  These top-$20$ results are based on histograms, which we can privatize using classical differentially private algorithms \cite{DworkMcNiSm06, McSherryTa07} or algorithms developed by our team \cite{DurfeeRo19}. These are the same algorithms we are currently using in our privacy system for the Audience Engagement API \cite{RogersSuPeDuLeKaSaAh20}. 

To determine which algorithm to select, we need to check some conditions on the dataset.  First, we need to determine whether the domain of the histogram of counts for each metric is known or unknown.  When the domain is unknown, the mere presence of a data element can leak privacy, regardless of the noise added to its count. The unknown domain algorithms take a parameter $\bar{d}$, which is the number of distinct elements (or rows) to fetch from the original dataset.  Second, we need to know if the number of distinct bins in the histogram that can be modified by a user is bounded or not (restricted or unrestricted sensitivity, also known as $\ell_0$-sensitivity).  If the histogram has restricted sensitivity, we denote $\Delta>0$ as the number of distinct bins that can change by adding or removing one user. Since we are working with event-level privacy we will use $\Delta = 1$, for top employers and top jobs.   All reports that we want to privatize will be based on distinct counts, so that a user can contribute at most 1 to the height of any single bin (also known as $\ell_\infty$-sensitivity) in the histogram that is used in each metric.  The table of algorithms is given in Table~\ref{table:tasks} and more detail on each algorithm can be found in \cite{RogersSuPeDuLeKaSaAh20}.

\begin{table}[htbp]
\centering\setcellgapes{4pt}\makegapedcells
\begin{tabular}{ |c|c|c| } 
 \hline
 & $\Delta$-restricted sensitivity & unrestricted sensitivity \\ 
 \hline
 \shortstack{Known \\ domain} & $\knownLap{\Delta}$ \cite{DworkMcNiSm06} & $\knownEM{k}$ \cite{McSherryTa07} \\ 
 \hline
\shortstack{Unknown \\ domain} & $\rTE{\Delta,\bar{d}}$ \cite{DurfeeRo19} & $\rT{k,\bar{d}}$ \cite{DurfeeRo19} \\ 
 \hline
\end{tabular}
\caption{DP algorithms for top-$k$ distinct count queries \label{table:tasks}}
\end{table}

One of the main benefits of the unknown domain algorithms is that they do not require knowing the set of elements to search for in advance.  Instead the data is used to determine which elements to publish, thus they can be used for top jobs or skills with no change to the algorithm. An additional benefit is that we know the results returned are actually in the dataset, meaning that top skills will only be shown if they are actually skills in the dataset that users added to their profile.  To ensure privacy, this means that we may not publish all the results, due to low counts, but we will not show a top job hiring unless that job actually hired users.  The known domain algorithms would require populating the dataset with all possible jobs, even if no users were hired at some jobs and thus noisy results could show one of these jobs as a top hiring job.  We opted for using the unknown domain algorithms so that elements whose true count was zero would not be shown.  For more detailed information about the algorithms, see \cite{RogersSuPeDuLeKaSaAh20}

%% file: privReports.tex

\section{Privatized Labor Market Insights}

We now cover the assumptions and private algorithms used in the three metrics described in the introduction.  Note that we provide reports for different dates, which aggregate the past three months' hires from the given date.  Hence, the report for May contains aggregated data for May, April, and March.  We also slice by countries, individual regions in each country, and industry.  Some reports might state on the webpage that there is insufficient data, this is due to privacy restrictions that we will now cover for each of the three metrics.

The focus below is on privacy loss parameters for differential privacy.  Note that we have incorporated additional privacy safeguards in our approach. We minimize the amount of data we publish by only showing top-20 results in each report. We  also make sure we provide percentages (where noise is also added to the denominators) and orderings of the DP results instead of noisy counts.

\subsection{Who is hiring?}

This metric is based on a histogram that gives the number of distinct hires for each employer using the last three months of data. There is a histogram for each month, each country/region, and each industry.  For each report date, country/region, and industry, we use the unknown domain and $(\Delta = 1)$-restricted sensitivity algorithm $\rTE{\Delta,\bar{d}}$.  In this algorithm, we first fetch the top-$(\bar{d} = 1000)$  employers hiring from the current 3 month period, add Laplace noise to the counts and, from the noisy counts, only show the top-20 results (top-10 on the webpage and top-20 available to download as CSV) with noisy counts above a noisy threshold.  In the algorithm, we add Laplace noise to all the counts in the top-$1000$ that has standard deviation $\sqrt{2}/0.6 \approx 2.36$ and only show elements if their noisy count is above a (noisy) threshold of approximately $40$.  

\begin{figure}[h]
\centering
\includegraphics[width=\columnwidth]{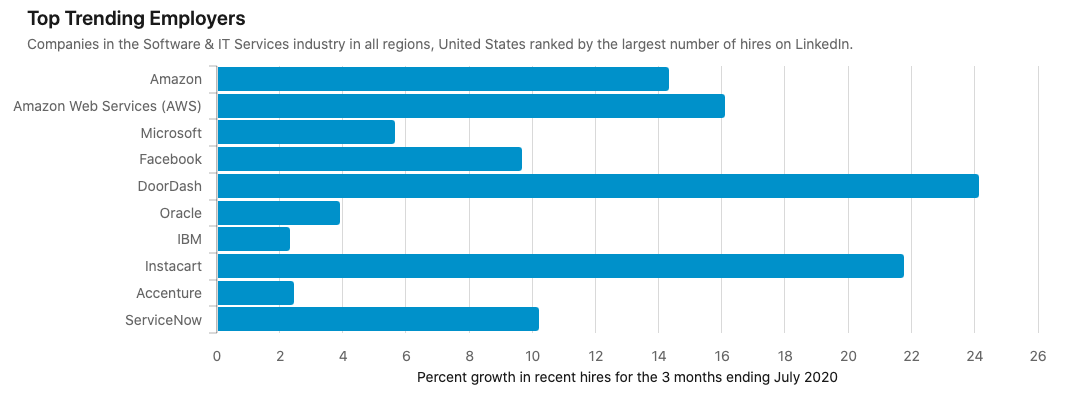}
\caption{Top employers in Software \& IT Services hiring in the U.S. for the July 2020 report.  The results are sorted from top to bottom by the outcome of $\rTE{\Delta,\bar{d}}$.  We also give the percent growth in recent hires based on noisy counts.\label{fig:top-hires}}
\end{figure}

Given this sorted list of private top-$20$ employers and their noisy counts, we then compute the number of hires made in the previous 3 month window for each of the top-$20$.  Because we have the set of employers already, we can then use the classical $\knownLap{\Delta}$, again with $\Delta = 1$, to return the noisy counts with the same scale of noise as in the top-$20$ calculation.  The numbers in Figure~\ref{fig:top-hires} are the ratios of the 3 months of noisy counts multiplied by 100, including the given date, and the previous 3 months' noisy counts, although they are sorted by the ranked list returned by $\rTE{\Delta,\bar{d}}$.

In the differential privacy analysis, we assume that one user cannot be hired to a new employer more than once in a three month period. This assumption holds for at least 95\% of the users in the dataset. Every report date, a hire will appear in 4 reports: the country report, the region report, the country-industry report, and the region-industry report. We chose parameters such that $\rTE{\Delta,\bar{d}}$ is $(0.6,10^{-10})$-DP and $\knownLap{\Delta}$ is $0.6$-DP. Therefore each report is $(1.2,10^{-10})$-DP because the ratio that uses both counts accumulates the privacy loss. Therefore, for each report date, we guarantee the top-employer reports are together $(4.8, 4 \times 10^{-10})$-DP.

\subsection{What jobs are available?}

This metric is based on a histogram that gives the number of distinct hires for each job using the last three months of data. There is a histogram for each report date, each country/region, and each industry.  Similar to what we computed for top employers, we use $\rTE{\Delta,\bar{d}}$ with $(\Delta = 1)$-restricted sensitivity and fetch the top-$(\bar{d}=1000)$ jobs hiring from the current three month window and only show the top-20 results above a threshold.  We use the same scale of noise as in the top employers hiring query.  We then divide each count by the total number of hires in the selected 3 month window date and in the selected country/region, which we privatize with Laplace noise with the same scale of noise.  We then multiply by 100 to get a percentage increase.

\begin{figure}[h]
\centering
\includegraphics[width=\columnwidth]{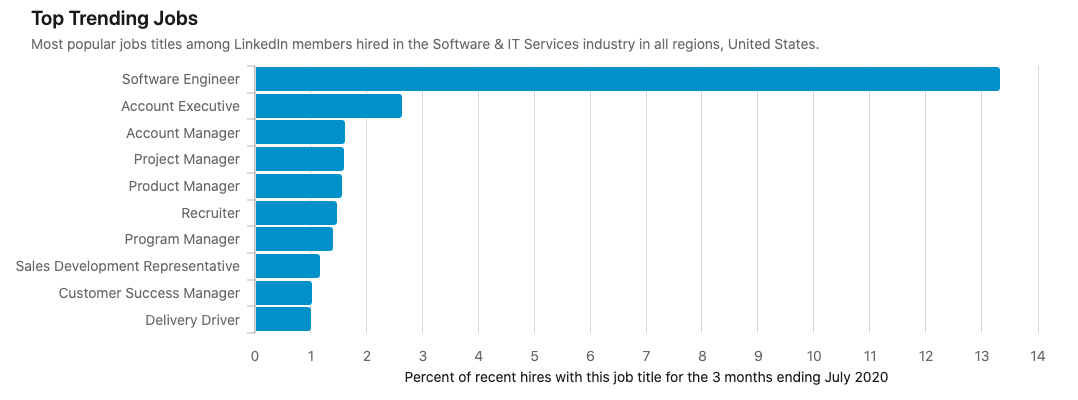}
\caption{Top jobs in Software \& IT Services hiring in the U.S. for the July 2020 report.  The results are sorted from top to bottom by the outcome of $\rTE{\Delta,\bar{d}}$.  We also give the percentage of recent hires associated with each job title.\label{fig:top-jobs}}
\end{figure}

The numbers in Figure~\ref{fig:top-jobs} are the ratios of the current month's noisy counts, which is the result of $\rTE{\Delta,\bar{d}}$, and the current month's total number of hires in a selected country/region, which has Laplace noise added to the true count.

In the differential privacy analysis, we assume a user gets hired to at most 1 occupation in a three month window (which is true for at least 95\% of the users in this dataset). Using the same algorithms as in the previous section, we ensure that each report is $(1.2,10^{-10})$-differentially private. Similarly, we guarantee that each month, the top-jobs query is $(4.8, 4 \times 10^{-10})$-DP.

\subsection{What skills are needed?} 

Based on the top trending jobs that we found in the previous query, we also want to present the top skills from those jobs that are hiring the most.  This will help users identify skills that they can learn to help get hired in the current economy.  To do this, we start with finding tuples of (country/region, job, skill) for each of the top jobs we previously computed for a given date.\footnote{A preprocessing step that uses TF-IDF was applied to clean the data. This was necessary because otherwise common skills like Microsoft Word would dominate. This preprocessing step was not done with DP algorithms and thus we can not provide an overall DP guarantee for this report type. Regardless, we applied our DP algorithms to the final histogram of skill counts to provide some uncertainty to the user’s true data. We also point out that each report date will have the same counts for skills if the same top jobs are found in different industries.  }  To determine the count of each skill in a particular country/region, we count the number of distinct users over a 5 year period with the same (country/region, job, skill) tuple and then take a max over all jobs to get the count of that skill in the country/region.  One user can have multiple skills in a 5 year period.  Hence, we put no bound on how many skills a single user can change the count of and treat the histogram in a report as if it has unrestricted sensitivity.  

\begin{figure}[h]
\centering
\includegraphics[width=\columnwidth]{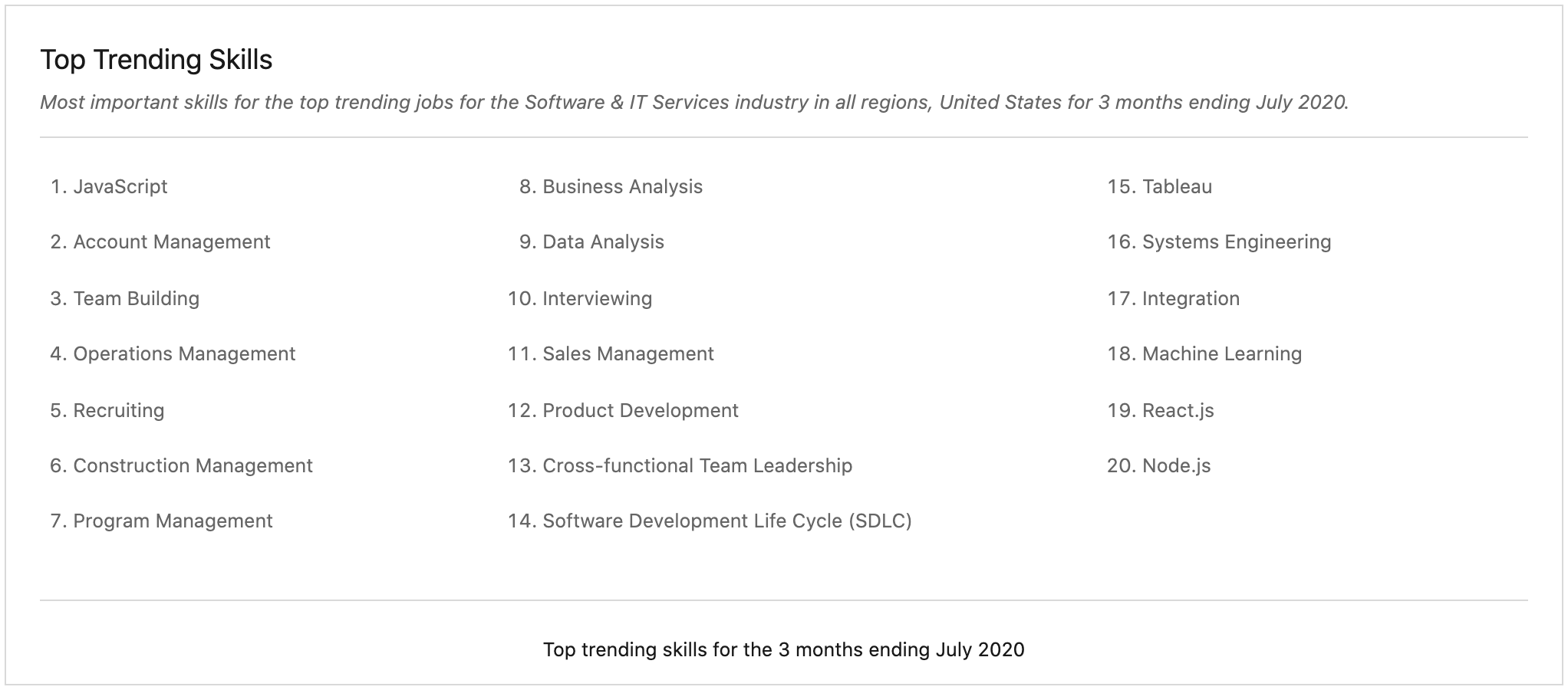}
\caption{Top skills from the top jobs in Software \& IT Services hiring in the U.S. for the July 2020 report in descending order from $\rT{k,\bar{d}}$.\label{fig:top-skills}}
\end{figure}

Since the top-jobs hiring report is already privatized, we need only account for the additional privacy leakage from publishing the skills from those jobs.  We then use $\rT{k,\bar{d}}$ with privacy parameters $\epsilon =0 .1, \delta= 10^{-10}$ where we fetch the skills with the top-$(\bar{d} = 1000)$ counts and only report the top-$(k=20)$.  Note that this algorithm is based on Gumbel noise, which allows us to select elements in the top-$k$, but not release their counts, despite having noise added.  Hence, we only return the ranked list of elements after adding Gumbel noise with scale parameter $10$ to the true skill counts and only returning an element if the noisy count is above a (noisy) threshold of roughly 260. See Figure~\ref{fig:top-skills} for the ranked top skills from $\rT{k,\bar{d}}$.  

%% file: conclusion.tex

\section{Conclusion}

We have detailed our approach for ensuring privacy of our members while also publishing timely data that can help governments, policy makers, and members of the global workforce make informed policy decisions during these challenging times.  We point out that it is important to continue advancing privacy technologies to maintain a balance between usability of data and protection of our member's privacy.  The trust of our members remains paramount in each of our products and we want to continue developing and deploying scalable privacy systems.  